\documentclass[submission,
copyright,
creativecommons
]{eptcs}
\providecommand{\event}{G.Ciobanu (Ed.):\\
MeCBIC 2012
\\
EPTCS 
Proceedings.  2012.
}
\usepackage{graphicx, amssymb}
\usepackage{wrapfig,subfig}
\usepackage{amsmath}
\usepackage{amsthm}
\usepackage{mathpartir}
\usepackage{verbatim}
\usepackage{color}
\usepackage{timestamp}
\usepackage{fancyvrb}
\usepackage[normalem]{ulem} 
\newcommand{\SALTA} [1] {}

\newcommand{\syntaxdef}{\mathrel{::=}}
\newcommand{\ou}{\;\;  \mid \;\; }

\newcommand{\these}{\vdash}

\newcommand{\FV}{\texttt{FV}}

\newcommand{\fns}{\texttt{fn}}

\newcommand{\parop}{\mid}
\newcommand{\nil}{\boldsymbol {0}}
\def\new#1{\scop{#1}}

\newcommand{\loc}[2]{ \{ #1 \}_{#2} }

\def\pep#1{#1.}
\def\scop#1{\pep{(\nu #1)}}

\newtheorem{mydef}{Definition}

\newcommand{\rulename} [1] {{\sc #1}}

\newcommand{\mov}{\texttt{mov}}

\newcommand{\shape}[1]{\texttt{space}(#1)} 
                                \newcommand{\sha}[1]{\texttt{shape}(#1)}
\newcommand{\dis}[2]{\texttt{dis}(#1,#2)}

\newcommand{\delay}{\texttt{delay}}

\newcommand{\ts}[2]{  \{ \!\!\{ #1 \}\!\!\}^{#2} }
\newcommand{\stoc}{{\it st}}

\newcommand{\nstoc}{{\it mv}}
\newcommand{\Redex}[1]{\Re_{#1}}
\newcommand{\Ctx}[1]{C_{#1}}
\newcommand{\hole}{[\;]}

\newcommand{\OK}{{\rm OK}}

\newcommand{\CtxInt}{C}
\newcommand{\redDet}[1]{\rightarrowtail}
\newcommand{\redStoc}[1]{\dashrightarrow}
\newcommand{\redTime}[1]{\stackrel{#1}{\rightsquigarrow}}
\newcommand{\redT}[1]{\labRed{#1}}
\newcommand{\labRed}[1]{{\xrightarrow{#1}}}
\newcommand{\stTime}{\tau}

\newcommand{\g}{\xi,\omega, \sigma}
\newcommand{\gP}{\xi',\omega', \sigma'}

\newcommand{\PI}{Pi}
\newcommand{\locP}[3]{\{#1\}_{#3}}

\definecolor{brown}{rgb}{0.85,.66,0}

\newif\ifsubmit
\submitfalse
\ifsubmit

\newcommand{\ACComm}[1]{}

\newcommand{\MDComm}[1]{}

\newcommand{\PGComm}[1]{}

\newcommand{\ATComm}[1]{}

\newcommand{\mdc}[1]{}

\newcommand{\acc}[1]{}

\newcommand{\atc}[1]{}
\else

\newcommand{\ACComm}[1]{{\scriptsize \textcolor{blue}{[Adriana{:} #1]}}}

\newcommand{\MDComm}[1]{{\scriptsize \textcolor{red}{[Mariangiola{:} #1]}}}

\newcommand{\PGComm}[1]{{\scriptsize \textcolor{magenta}{[Paola{:} #1 ]}}}

\newcommand{\ATComm}[1]{{\scriptsize \textcolor{brown}{[Angelo{:} #1 ]}}}

\newcommand{\mdc}[1]{\textcolor{red}{[Mariangiola{:} #1]}}

\newcommand{\acc}[1]{\textcolor{blue}{[Adriana{:} #1]}}

\newcommand{\atc}[1]{\textcolor{brown}{[Angelo{:} #1]}}
\fi

\makeatletter
 \let\@copyrightspace\relax
 \makeatother

\title{Parallel BioScape: A Stochastic and Parallel Language for Mobile and Spatial Interactions \thanks{Extended Abstract. This research is funded by the BioBITs Project (\emph{Converging
Technologies} 2007, area: Biotechnology--ICT), Regione Piemonte.}}
\author{Adriana Compagnoni
\institute{Department of Computer Science\\Stevens Institute of Technology}
\and Mariangiola Dezani--Ciancaglini
\institute{Dipartimento di Informatica\\Universit\`a di Torino}
\and Paola Giannini
\institute{Dipartimento di Informatica\\Universit\`a del Piemonte Orientale}
\and Karin Sauer
\institute{Department of Biological Sciences\\Binghamton University}
\and Vishakha Sharma
\institute{Department of Computer Science\\Stevens Institute of Technology}
\and Angelo Troina
\institute{Dipartimento di Informatica\\Universit\`a di Torino}
}

\def\copyrightholders{A.\ Compagnoni, M.\ Dezani--Ciancaglini, P.\ Giannini,\\
  K.\ Sauer, V.\ Sharma, A.\ Troina}

\def\titlerunning{Parallel BioScape: A Stochastic and Parallel Language for Mobile and Spatial Interactions}

\def\authorrunning{A.\ Compagnoni, M.\ Dezani--Ciancaglini, P.\ Giannini,
  K.\ Sauer, V.\ Sharma, A.\ Troina}
\begin{document}
\maketitle

\begin{abstract}BioScape is a concurrent language motivated by the biological
landscapes found at the interface of biology and biomaterials
\cite{BioScape:CS2Bio}. It has been motivated by the need to model
antibacterial surfaces, biofilm formation, and the effect of DNAse in
treating and preventing biofilm infections.
As its predecessor, SPiM
\cite{Phillips07}, BioScape has a sequential semantics based on Gillespie's
algorithm \cite{Gillespie_1977}, and its
implementation does not scale beyond 1000 agents. However, in order to
model larger and more realistic systems, a semantics that may take advantage
of the new multi-core and GPU architectures is needed. This motivates the
introduction of parallel semantics, which is the contribution of this paper: Parallel
BioScape, an extension with fully parallel semantics.
\end{abstract}

Process algebras have been successfully used in the modeling of
biological systems, see \cite{Priami01, CardelliGTP09, Baoetal2010},
where they are
particularly attractive, because of their ability to accommodate new
objects and new behavioral attributes as the complex biological system
becomes better understood.
However, most of the modeling languages lack adequate support for the design of
systems in which to study complex interactions involving both spatial
properties, movements in three-dimensional space, and stochastic interactions.
Recently, new spatial modeling languages
allowing explicit description of temporal spatial dynamics of
biochemical processes have been proposed (SpacePi \cite{spacePi},
DCA \cite{DCA}, \textit{L}$\Pi$ \cite{lpi}, Stochsim
\cite{LeNovereEtAl:2001}).
Other agent-based platforms \cite{ABMS}
include
C-Immsim \cite{ISSO, CImmSim} and PathSim visualizer
\cite{PathSim}.
However, few of them support
individual based, continuous motion, and continuous space stochastic
simulation \cite{Bittig10}, which are important features
for modeling temporal spatial dynamics of biochemical processes
accurately.  To address this problem in previous work we introduced
BioScape \cite{BioScape:CS2Bio},
a  language incorporating both stochasticity and 3D spatial attributes.


Gillespie's algorithm produces two outputs in each iteration: 1) the
next reaction \textit R to be executed and 2) a slice of time \textit
t to advance the simulation clock. Since many reactions, including
many instances of the same reaction, may be available, the slice of
time \textit t does not correspond to the time that \textit R would
take, but an amount of time proportional to the time it would take to
execute all available reactions. In contrast, the parallel semantics
will execute all available reactions, not just one instance of one
reaction \textit R, and the first challenge is then how to calculate simulated
time.
Reaction times can vary substantially, for example, some
prokaryotic cell mitosis takes ten minutes, some plant cell mitosis
takes about half an hour, while some animal cell mitosis takes about
three hours.
If we trigger all reactions together, how do we advance the
simulation clock?  The solution we propose here consists of annotating
each product of a reaction with a timer indicating how long that
reaction will take.

For example, if $Cell \rightarrow_{30} Cell  \parop  Cell$ means that a
$Cell$ takes 30 minutes to split, through mitosis, into two daughter cells, then we will
annotate the two daughter cells as $\ts{Cell}{30}$ and
$\ts{Cell}{30}$.
As time lapses, the timer will be reduced, and when reaching $\ts{Cell}{0}$,
both cells will be available for new reactions.

In Fig.~\ref{fig.syntax} we define the syntax of the calculus, which
slightly simplifies the syntax of \cite{BioScape:CS2Bio} in order to
avoid  decorating semantic  processes with shapes, as defined at page \pageref{shape}.

\begin{figure}[t]
\vspace*{-0.8cm}
 {\small \begin{align*}
          P,Q & \syntaxdef  \quad \nil \quad \ou X(\overline{u}) \quad
          \ou P \parop Q\quad \ou \new{a@\texttt{r},\texttt{rad}}P &    D & \syntaxdef \quad \emptyset \quad \ou D,X(\overline{x}) = M^{\g}\quad\FV(M)\subseteq\overline{x} \\
          M &\syntaxdef   \quad \pi.P \; [+\; M] &           u,v & \syntaxdef \quad a\ou b\ou\cdots \ou x\ou y\ou \cdots\\
          \pi &  \syntaxdef  \quad \delay @\texttt{r} \quad  \ou !u(v) \quad  \ou ?u(x)  \quad \ou  \mov &
          E & \syntaxdef \quad \emptyset \quad \ou E,
          {a@\texttt{r},\texttt{rad}}
\end{align*} }
\vspace*{-0.6cm}
\caption{Syntax}
 \label{fig.syntax}
\vspace*{-0.5cm}
\end{figure}

We assume a set of channel names, denoted by
$a$, $b$, and a set of variables, denoted by $x$,
$y$, with subscripts or superscripts, if needed. As usual,
$\overline{a}$ is $a_1,\ldots,a_n$, and similar for
$\overline{x}$.
The empty process is $\nil$.
By $X(\overline{u})$ we denote an instance of the entity defined by $X$.
The actual parameters of the instance may be either channel names or variables,
in case the instance occurs in a definition.
The process $ P \parop Q$ is the parallel composition
of processes $ P$ and  $Q$.
 By
$\new{a@\texttt{r},\texttt{rad}}P$ we
define the channel name $a$ with two parameters \texttt{r} and
\texttt{rad}$\in \mathbb{R}_{\geq0}$ within process $P$; the parameter \texttt{r} is the stochastic
rate for communications through channel $a$ and $\texttt{rad}$ is the
communication radius. The radius is the maximum distance between
processes in order to communicate through channel $a$, and the
reaction rate determines how long it takes for two processes to react given that they are close enough
to communicate.

The \textit{heterogeneous} choice is denoted by $M$,   where $\pi.P \; [+\;
M]$ means $\pi.P \ou \pi.P \;+\;
M$.    Choices may
  have reaction branches and movement branches.  The reaction branches
  are probabilistic (stochastic), since reactions are subject
  to kinetic reaction rates, while the movement branches are
  non-deterministic, since the movement of instances of entities is always
  enabled, provided there is enough space. The prefix $\pi$ denotes the action that the process
  $\pi.P$ can perform. The prefix $\delay @\texttt{r}$ is a
  spontaneous and unilateral reaction of a single process, where
  \texttt{r} is the stochastic rate. The prefix $!u$ denotes output,
  and the prefix $?u$ denotes input.  The prefix $\mov$ moves
    processes in space according to their diffusion rate ($\omega$) (see below).
We use standard syntactic
abbreviations such as $\pi$ for
$\pi.0$.

We denote by $D$ a global list of definitions. The equality $X(\overline{x})= M^{\g}$
defines entity $X$ with formal parameters $\overline{x}$, to be 
the choice $M$ with geometry
$\g$, specifying a movement space $\xi$, a step
$\omega$, and a shape $\sigma$. The choice $M$ describes the behavior of $X$
  with a choice of prefixed processes. The selection of one of the
  choices depends not only on the available interactions with other
  processes, but also on the available space.
The movement space $\xi$ is a set of
point coordinates in the global coordinate system defining a
volume. Intuitively, $X$ can move within $\xi$.  The step $\omega \in
\mathbb{R}_{\geq0}$, is the distance that $X$ can stir in a movement,
and it corresponds to the diffusion rate of $X$;
$\sigma$ is the three-dimensional shape (sphere, cube, etc.) of $X$.
The movement space for the empty process $\nil$
is everywhere, the global space, and its movement step is 0 by
default.  The entity variable $X$ can be defined at most once  in $D$, and the free variables of
$P$, denoted by $\FV(P)$, must be a subset of the variables
$\overline{x}$.
We also write $X(\overline{x})= (\pi.\pi'.P)^{\g}$ as short for $X(\overline{x})= (\pi.Y(\overline{x}))^{\g}$ and $Y(\overline{x})= (\pi'.P)^{\g}$.

We use $E$ to range over environments of channel name
declarations. By $a@\texttt{r},\texttt{rad}$ we declare channel name $a$
with reaction rate \texttt{r} and reaction radius \texttt{rad}. A
channel name $a$ appears at most once in $E$.

Consider the following simple example of a bacterium \texttt{Bac},
that can either move or divide into two daughter cells. \texttt{Bac}
is defined with movement space \texttt{movB}, movement step
\texttt{stepB}, and shape \texttt{shapeB}. Intuitively, bacteria can
move within \texttt{movB}, with non-deterministic steps of length \texttt{stepB},
and the shape \texttt{shapeB} is at all times contained within
\texttt{movB}.  The prefix \texttt{ mov } represents a non-deterministic movement
of length \texttt{stepB}, whereas \texttt {delay@1.0.(Bac() | Bac())}
represents mitosis, the division of a bacterium into two daughter
cells: \texttt{Bac() | Bac()}, and the \texttt{delay@1.0} prefix is
used to model the fact that division is not an instantaneous reaction.\\
\centerline{
${\tt Bac()} = ( {\tt  mov.Bac() \; +} \; {\tt delay@1.0.(Bac() | Bac())})^{\tt movB,stepB,shapeB}$
}
\begin{figure}[tbp]
\footnotesize{
    \begin{mathpar}
      \inferrule[S.Loc]
      { P \equiv Q}
      { \locP{P}{\sigma}{\mu} \equiv \locP{Q}{\sigma}{\mu} }
      \and
     \inferrule[S.Loc.Nu]
       { }
       { (\nu a@\texttt{r}, \texttt{rad}).\locP{P}{\sigma}{\mu}\equiv \locP{(\nu a@\texttt{r}, \texttt{rad}).P}{\sigma}{\mu}}
        \and
      \inferrule[S.Loc.Par]
      { \mu_1(\sha{P})\cup\mu_2(\sha{Q})=\mu(\sha{P\parop Q})}
      { \locP{P}{\sigma}{\mu_1} \parop \locP{Q}{\sigma}{\mu_2} \equiv \locP{P\parop Q}{\sigma}{\mu} }
     \and
     \inferrule[S.Nu.Com]
       { }
       { (\nu a@\texttt{r}, \texttt{rad}).(\nu b@\texttt{r}', \texttt{rad}').A\equiv (\nu b@\texttt{r}', \texttt{rad}').(\nu a@\texttt{r}, \texttt{rad}).A}
  \and
    \inferrule[S.Nu.Par]
       { a\not\in\fns(B)}
       { ((\nu a@\texttt{r}, \texttt{rad}).A)\parop B\equiv (\nu a@\texttt{r}, \texttt{rad}).(A\parop B)}
    \end{mathpar}}
\vspace*{-0.6cm}
  \caption{Structural Equivalence of Spatial Configurations}
\vspace*{-0.5cm}
  \label{fig.pi.struct}
\end{figure}
A run-time system is represented by a parallel composition of entity instances (without free variables) each
  with its shape, 
  and located in some positions of a global frame.
We define the {\texttt shape} of processes inductively as follows:\\
  $\begin{array}{lll}
\sha{\nil}  = \emptyset&\qquad&
\sha{X(\overline{a})}  = \sigma~~\mbox{if}\ X(\overline{x})= M^{\xi,\omega,\sigma}\in D\\
\sha{\new{a@\texttt{r},\texttt{rad}}P} = \sha{P}&&
\sha{P \parop Q}  = \sha{P} \Cup \sha{Q}\\
\end{array}$\\
where $\Cup$ gives a shape obtained by composing two shapes trough juxtaposition. For different applications we can choose suitable functions to realise $\Cup$, we only require $\Cup$ to be a commutative and associative operator, i.e. $\sigma_1\Cup\sigma_2=\sigma_2\Cup\sigma_1$ and $(\sigma_1\Cup\sigma_2)\Cup\sigma_3=\sigma_1\Cup(\sigma_2\Cup\sigma_3)$.

%
%

  We use $\mu$ to denote a map which applied to a shape locates it in the global space, by putting its
barycentre at a fixed point, orienting the shape, and possibly modifying it.
So $\mu(\sha{P})$ computes the space occupied by a
  process $P$ in the global coordinate system.\label{shape}
  Processes may also share channels
  for communication. {\em Spatial configurations}, denoted by $A$,
  $B$, $\ldots$ are defined as follows:\\
\centerline{$
  A,B ~  \syntaxdef \locP{P}{\sigma}{\mu} \ou A \parop B \ou \
  \new{a@ \texttt{r}, \texttt{rad}}. A
$
}
Structural equivalence on configurations is defined in Fig.~\ref{fig.pi.struct}, omitting the rules for associativity and commutativity of $\parop$ and
$+$. Parallel composition  has
neutral element $\locP{0}{\sigma}{\mu}$ for any $\mu$.
Rule \rulename{S.Loc} uses
the standard structural equivalence of \PI-calculus processes.
The premise
  of rule \rulename{S.Loc.Par} assures that the two equivalent
  processes occupy exactly the same space.
  In rule
\rulename{S.Nu.Par}, $\fns$ is a function that returns the set of free channel names of a configuration.
\begin{figure}[bp]
\vspace*{-0.4cm}
\footnotesize{
    \begin{mathpar}
     \inferrule[SR.Delay]
        {X(\overline{x})= (delay@\texttt{r}.P\;[+\;M])^{\xi,\omega,\sigma}\in D }
        {E\these \locP{X(\overline{a})}{\sigma}{\mu}\labRed{\texttt{r}} \locP{P[\overline{a}/\overline{x}]}{\sigma}{\mu}}\and
        \inferrule[SR.Str]
        { A \equiv A'\\ E\these A' \labRed{\texttt{r}} B'\\ B' \equiv B }
        { E\these A \labRed{\texttt{r}} B }\and
  \inferrule[SR.Com]
        {X(\overline{x})= (!a(b).P\;[+\;M])^{\xi,\omega,\sigma}\in D\quad
        Y(\overline{y})= (?a(z).Q\;[+\;N])^{\xi',\omega',\sigma'}\in D\quad \dis{\mu} {\mu'} \leq\texttt{rad}}
        {E,   a@\texttt{r}, \texttt{rad}\these  \locP{X(\overline{c})}{\sigma}{\mu} \parop
          \locP{Y(\overline{d})}{\sigma'}{\mu'}
          \labRed{\texttt{r}}\locP{P[\overline{c}/\overline{x}]}{\sigma}{\mu} \parop
          \locP{Q[\overline{d}/\overline{y}][b/z]}{\sigma'}{\mu'} }
%
    \end{mathpar}}
\vspace*{-0.6cm}
  \caption{Stochastic Reduction Relation}
\vspace*{-0.5cm}
  \label{fig:stoc.reduction}
\end{figure}

\begin{figure}[tbp]
\footnotesize{
    \begin{mathpar}
  \inferrule[NR.Move]
      { \mu'=\texttt{translate}(\omega,\mu)  \quad
         \mu'(\sigma) \subseteq         \xi \quad
        X(\overline{x})= (\mov.P \;[+\;M])^{\xi,\omega,\sigma}\in D}
      { \locP{X(\overline{a})}{\sigma'}{\mu} \labRed{}
        \locP{P[\overline{a}/\overline{x}]}{\sigma'}{\mu'}}
     \and
        \inferrule[NR.Str]
        { A \equiv A'\\  A' \labRed{} B'\\ B' \equiv B }
        {  A \labRed{} B }
    \end{mathpar}}
  \vspace*{-0.6cm}
  \caption{Non-stochastic Reduction Relation}
\vspace*{-0.5cm}
  \label{fig.rednorate}
\end{figure}

The (parallel) operational semantics of BioScape is based on two {\em
  auxiliary} 
  reduction
relations: a stochastic
relation, $E\these A \labRed{\texttt{r}} B$, for reactions such as
synchronisation and delay, defined in Fig.~\ref{fig:stoc.reduction}, and
a non-deterministic (non-stochastic) relation, $A \labRed{} B$, for
geometric transformations, in our case movement, defined in
Fig.~\ref{fig.rednorate}. Notice that reduction axioms (\rulename{SR.Delay, SR.Com, NR.Move}) only involve entities ($X(\overline{a
})$), and entities evolve according to one of the choices  in their
definitions. In rules \rulename{SR.Delay, SR.Com} and
\rulename{NR.Move},  there is no  check of whether the entities of the
resulting process have enough space, since this check is done in the
parallel reductionrules \rulename{PR.Stoc}, and \rulename{PR.Move} of
Fig.~\ref{fig:par.reduction}. In particular, a stochastic
(non-stochastic) redex is stuck, if there is not enough space for its
reduct in the configuration. Therefore, the {\em evolution of systems in parallel BioScape  produces configurations in which space is consistent}.

Fig.~\ref{fig:stoc.reduction} defines the stochastic reduction
relation of BioScape, $E\these A \labRed{\texttt{r}} B$, where
$\texttt{r}$ is the rate of the channel used for synchronization or delay.
 We write $\dis{\mu}{\mu'}$
for the distance between the origin of
$\mu$ and the origin of ${\mu'}$. In rule
\rulename{SR.Com} the condition $\dis{\mu}{\mu'} \leq \texttt{rad}$
ensures that located processes
$\locP{X(\overline{c})}{\sigma}{\mu}$ and $\locP{Y(\overline{d})}{\sigma'}{\mu'}$ are close enough to communicate through
channel $a$.
The non-stochastic reduction relation  of BioScape,
 $ A \labRed{} B$, is defined in Fig.~\ref{fig.rednorate}.
By  \texttt{translate}($\omega$,$\mu$) we denote the function that randomly generates a new
map ${\mu'}$, using the movement step $\omega$ and the old
map $\mu$. The condition ${\mu'}(\sigma) \subseteq \xi$ of rule \rulename{NR.Move}
ensures the new located process
$\loc{P[\overline{a}/\overline{x}]}{\mu'}$ is within 
the movement space $\xi$ of $X$ (see previous remark about not checking if the entity moves to an empty space).

For stochastic reductions we compute the duration of the reduction, based on the exponential distribution associated with the propensity of the reduction. Since reductions may have different durations, we introduce {\em timed configurations}, $\ts{A}{n}$, meaning that, after a time $n$, this configuration will be $A$. We extend structural equivalence to timed configurations by adding that $\ts{A}{0}\equiv A$, and $A \equiv B$ implies $\ts{A}{n}\equiv \ts{B}{n}$. With the metavariables $F$, and $G$ we denote either spatial configurations or timed configurations ({\em extended configurations}), i.e.,\\
\centerline{
 $F,G ::=  A \ou \ \ts{A}{n}\ou F\parop G\ou\new{a@\texttt{r},\texttt{rad}}F\quad\quad (n\ge 0)$
}
%
%

We define a reduction strategy that given the whole configuration,
first moves all the processes that can be moved, and then executes all
the stochastic reductions that can be executed, omitting only
reductions which would lead to overlaps,
i.e. configurations where some entities occupy the same space. Both
non-stochastic and stochastic reductions are applied in parallel. For
this purpose, we define multi-hole contexts $\Ctx{}$ by the following
grammar:\\
\centerline{
$\Ctx{} \syntaxdef  F 
 \ou \hole \ou \Ctx{}\parop \Ctx{} \ou
  \new{x@r, \texttt{rad}} \Ctx{}$
}
Congruence on multi-hole contexts is naturally induced by the congruence on configuration, associativity and commutativity of the parallel operator, and standard rules for $\nu$ restrictions similar to \rulename{S.Nu.Com} and \rulename{S.Nu.Par}. Given this congruence any multi-hole context, $\Ctx{}$, may be written in a {\it canonical form}. That is, there is $\Ctx{}'$, $\Ctx{}\equiv\Ctx{}'$ such that $\Ctx{}'=\nu_1.\ldots\nu_n.F_1\parop\cdots\parop F_m\parop\hole\parop\cdots\parop\hole$,
where $\nu_i$, $1\leq i\leq n$, is an abbreviation for $\nu a_i@{\tt r}_i, {\tt rad}_i$, and for all $j$, $1\leq j\leq m$, $F_j=\ts{A}{n}$ for some $A$, and $n$, or $F_j=\locP{P}{\sigma}{\mu}$ for some $P$, and $\mu$. We say that $a_1@{\tt r}_1, {\tt rad}_1, \ldots, a_n@{\tt r}_n, {\tt rad}_n$ is ${\tt restr}(C)$.
In the following we assume that multi-hole contexts are always in canonical form.

As already mentioned, our reduction strategy avoids spatial
overlaps. In particular for moving reductions we have to ensure
that moves and reshaping are compatible with the available space, that
is after moving no entity overlaps with another entity. For stochastic reductions we have to assure that the created entities have their space.
To this aim we define the space of a configuration, and a predicate that says whether a configuration does not have any overlapping entities.

Let $\shape{F}$ be a function on configuration $F$ that returns the
space occupied by its processes located in the global frame defined as follows.\\
\centerline{$\begin{array}{lllll}
\shape{\locP{P}{\sigma}{\mu}} & = \mu(\sha{P})&\qquad\qquad&\shape{\ts{A}{n}}& = \shape{A}
\\
\shape{F \parop G} & = \shape{F} \cup \shape{G}&\qquad\qquad&
\shape{\new{a@\texttt{r},\texttt{rad}}F} & = \shape{F}
\end{array}$}
We say that a configuration $F$ is $OK$ if the various entities in $F$ do not overlap, that is:\\
\centerline{$
\begin{array}{l}
\locP{P}{\sigma}{\mu}\ \OK\qquad\qquad
A\ \OK~\Rightarrow~\ts{A}{n}\ \OK\qquad\qquad
F\ \OK~\Rightarrow~\new{x@\texttt{r},\texttt{rad}}F\ \OK\\
F\ \OK\ \wedge\  G\ \OK\ \wedge\  \shape{F}\cap\shape{G}=\emptyset~\Rightarrow~F \parop G\ \OK
\end{array}$
}
With the notion of $\OK$ configuration we define  two notions of
{\it well-formedness of configurations}.
The first notion is to be used for parallel
move reductions and the second for parallel stochastic reductions.
Theses notions are to be used to enforce $(i)$ the fact that only
reductions that have enough space for their reduct are allowed, and
$(ii)$ that we want {\em maximal parallelism}, that is any ``extra''
movement or transformation would produce an overlap.  In order to formalise this
we first need to single out the sets $\Redex{\nstoc}$ and $\Redex{\stoc}$ of movement and stochastic redexes, i.e. we define:\vspace{-1mm}
\begin{itemize}
\item $\Redex{\nstoc}=\{\locP{X(\overline{a})}{\sigma}{\mu}\ou X(\overline{x})=(\mov.P+M)^{\g}\in D
\}$,
\item $\Redex{\stoc}=\{\locP{X(\overline{a})}{\sigma}{\mu}\ou X(\overline{x})=(delay@\texttt{r}.P+M)^{\g}\in D
\}\cup$\\
$\{\locP{X(\overline{c})}{\sigma}{\mu} \parop \locP{Y(\overline{d})}{g'}{\mu'}\ou X(\overline{x})=(!a(b).P +M)^{\g}\in D ~\&~Y(\overline{y})=(?a(z).Q+N)^{\gP}\in D$\\
$~\&~\dis{\mu} {\mu'} \leq\texttt{rad}\}$
 where $a@{\tt r}, {\tt rad}$ is the declaration of channel $a$.
%
\end{itemize}\vspace{-1mm}
We extend the syntax of configurations by allowing {\em underlined  extended configurations}, defined by: an underlined  extended configurations is a configuration in which some spatial sub-configurations may be underlined. Underlined configurations
are the tool we use to define maximal parallelism. We can then define:\vspace{-1mm}
\begin{mydef}
\begin{enumerate}
\item An extended configuration $F$ is $\OK_\nstoc$ if $F$ is $\OK$ and $F\equiv \Ctx{}[A]$ with $A$ not underlined and $A\in\Redex{\nstoc}$ and  $A\labRed{} B$ imply $\Ctx{}[B]$ not $\OK$.\vspace{-1mm}
\item An extended configuration $F$ is $\OK_\stoc$ if $F$ is $\OK$ and $F\equiv \Ctx{}[A]$ with $A$ not underlined and $A\in\Redex{\stoc}$ and  $A\labRed{\texttt{r}} B$ imply $\Ctx{}[B]$ not $\OK$.
\end{enumerate}
\end{mydef}\vspace{-1mm}
As a last notion, we say that a context $C$ is {\em untimed} if it does not contain timed configurations.

\begin{figure}[tbp]
\vspace*{-0.2cm}
\footnotesize{
    \begin{mathpar}
      \inferrule[PR.Move]
        { 
           F_i\labRed{}G_i \quad (1\leq i\leq p)
           \\
       \Ctx{}[\underline{G_1}]\cdots[\underline{G_p}]\ \OK_\nstoc \\
        }
        {\Ctx{}[F_1]\cdots[F_p]\redDet{}\Ctx{}[G_1]\cdots[G_p]}

    \inferrule[PR.Timed] {
      n=min\{n_i\;|\;1\leq i\leq p\}\quad \Ctx{}\text{ is untimed}}
      {\Ctx{}[\ts{A_1}{n_1}]\cdots[\ts{A_p}{n_p}]\redTime{}\Ctx{}[\ts{A_1}{n_1-n}]\cdots[\ts{A_p}{n_p-n}]
      }

      \inferrule[PR.Stoc]
      {
        {\tt restr}(\Ctx{})\these A_i\labRed{\texttt{r}_i} B_i\quad
         \quad n_i=\stTime(\texttt{r}_i,\CtxInt_i[A_i])
        \quad
        (1\leq i\leq p)\\
         \Ctx{}[\CtxInt_{1}[\underline{B_1}]]\cdots[\CtxInt_{p}[\underline{B_p}]]\ \OK_\stoc
      }
      {\Ctx{}[\CtxInt_1[A_1]]\cdots[\CtxInt_{p}[A_p]]\redStoc{}\Ctx{}[\CtxInt_1[\ts{B_1}{n_1}]]\cdots[\CtxInt_{p}[\ts{B_p}{n_p}]] }

\inferrule[PR.Conf]
{F\redDet{} F_1\redStoc{} F_2\redTime{}F'}
{F\redT{} F'}
    \end{mathpar}}
\vspace*{-0.6cm}
  \caption{Parallel Reduction Relation}
\vspace*{-0.5cm}
  \label{fig:par.reduction}
\end{figure}
We are now able to explain our  parallel reduction strategy, whose rules are given in Fig.~\ref{fig:par.reduction}. The first three rules deal respectively with parallel movements, timed reductions, and stochastic reductions, while the fourth rule maps extended configurations into extended configurations by applying first the parallel movements, then the stochastic interactions, and finally by advancing the time of the minimum required to complete one or more interactions. In this way at the next iteration there would be new entities to be moved and/or stochastically reduced.

The condition of obtaining an $\OK_\nstoc$ extended configuration in
rule  \rulename{PR.Move} assures that all possible moves in
$\Ctx{}[F_1]\cdots[F_p]$ which do not cause overlaps have been done in
the reduction.  Similar effect is produced by the conditions that the
extended configuration is $\OK_\stoc$ and that the context is timed in
the following two rules, respectively. Rule \rulename{PR.Stoc}
prescribes that the time of a stochastic reaction depends (through the
function $\stTime$) on the  rate of the reduction and on the number of
available reactants. The outer context $C$ is a multi-hole context, while the context $\Ctx{i}$ of the redex $A_i$ is a
single hole context capturing the surrounding environment that influences the speed of the reduction.
We could incorporate a counting function keeping track of the available reactants in the communication range (in a way similar to what is done, e.g., in~\cite{BDGT12,CWCTCS}).

Examples, results of
simulations, comparisons with related papers and discussions  can be found in the full version of this papers available at \url{http://www.di.unito.it/~dezani/papers/cdgsst.pdf}.
\vspace{-7mm}
\bibliographystyle{eptcs}
\bibliography{TPrefs}
\end{document}